\documentclass[sigconf]{acmart}

\usepackage{enumitem}
\AtBeginDocument{%
  }

\copyrightyear{2025}
\acmYear{2025}
\setcopyright{acmlicensed}
\acmConference[UMAP '25]{33rd ACM Conference on User Modeling, Adaptation and Personalization}{June 16--19, 2025}{New York City, NY, USA}
\acmBooktitle{33rd ACM Conference on User Modeling, Adaptation and Personalization (UMAP '25), June 16--19, 2025, New York City, NY, USA}
\acmDOI{10.1145/3699682.3728345}
\acmISBN{979-8-4007-1313-2/2025/06}

\begin{document}

\title[With Friends Like These, Who Needs Explanations?]{With Friends Like These, Who Needs Explanations? \\ Evaluating User Understanding of Group Recommendations}



\author{Cedric Waterschoot}
\affiliation{%
  \institution{Maastricht University}
  \city{Maastricht}
  \country{The Netherlands}}
\email{cedric.waterschoot@maastrichtuniversity.nl}

\author{Raciel Yera Toledo}
\affiliation{%
  \institution{University of Ja\'en}
  \city{Ja\'en}
  \country{Spain}}
\email{ryera@ujaen.es}

\author{Nava Tintarev}
\affiliation{%
  \institution{Maastricht University}
  \city{Maastricht}
  \country{The Netherlands}}
\email{n.tintarev@maastrichtuniversity.nl}

\author{Francesco Barile}
\affiliation{%
  \institution{Maastricht University}
  \city{Maastricht}
  \country{The Netherlands}}
\email{f.barile@maastrichtuniversity.nl}

\renewcommand{\shortauthors}{Waterschoot et al.}

\begin{abstract}

Group Recommender Systems (GRS) employing social choice-based aggregation strategies have previously been explored in terms of perceived consensus, fairness, and satisfaction. At the same time, the impact of textual explanations has been examined, but the results suggest a low effectiveness of these explanations. However, user understanding remains fairly unexplored, even if it can contribute positively to transparent GRS. This is particularly interesting to study in more complex or potentially unfair scenarios when user preferences diverge, such as in a minority scenario (where group members have similar preferences, except for a single member in a minority position). In this paper, we analyzed the impact of different types of explanations on user understanding of group recommendations. We present a randomized controlled trial ($n=271$) using two between-subject factors: (i) the aggregation strategy (additive, least misery, and approval voting), and (ii) the modality of explanation (no explanation, textual explanation, or multimodal explanation). We measured both subjective (self-perceived by the user) and objective understanding (performance on model simulation, counterfactuals and error detection). In line with recent findings on explanations for machine learning models, our results indicate that more detailed explanations, whether textual or multimodal, did not increase subjective or objective understanding. However, we did find a significant effect of aggregation strategies on both subjective and objective understanding. These results imply that when constructing GRS, practitioners need to consider that the choice of aggregation strategy can influence the understanding of users. Post-hoc analysis also suggests that there is value in analyzing performance on different tasks, rather than through a single aggregated metric of understanding.
\end{abstract}




\maketitle

\section{Introduction}
Group Recommender Systems (GRS) process the preferences of multiple individuals to derive a single recommendation tailored to suit the group as a whole. To achieve this, previous research has introduced social choice-based aggregation strategies \cite{masthoff2015group}. These strategies present a range of options in which the preferences of individual group members are aggregated and present an opportunity to adapt GRS to different group configurations and needs. 
However, the implementation of these strategies on top of already complex recommender systems may result in non-transparent models, hindering user understanding. This is particularly interesting in scenarios in which user preferences diverge, leading to potentially unfair scenarios. Social choice-based explanations for group recommendations have been used as a way to intuitively explain the process behind group recommendations \cite{najafian2018generating,kapcak2018tourexplain}. These textual explanations have been evaluated in terms of users' fairness perception, consensus perception, and satisfaction in user studies with contrasting results \cite{tran2019towards,barile2023evaluating}. Hence, their effectiveness in improving user understanding of the strategies used to generate recommendations for the group still remains to be seen. In this study, we build upon these works, focusing on the understandability of explanations and in turn, group recommendations themselves. We formulate the following research questions:\newline

\noindent \textbf{RQ1.} Do explanations increase users’ understanding of the underlying social choice-based aggregation strategies?

\noindent \textbf{RQ2.} Does the effect of explanations on the understandability of GRS vary depending on the underlying social-based aggregation strategy?\newline

Inspired by \citet{Wang2021}, we constructed a randomized controlled trial with two between-subject factors ($3x3$ design): (i) the social choice-based aggregation strategy used for generating group recommendations, and (ii) the type of explanation provided to the participant. We included three explanation types: \textit{no explanation} (control group), \textit{textual explanation only} and, \textit{a multimodal explanation} combining text with a visualization. To measure the participant's understanding of the strategy and GRS, we made use of \textit{objective} and \textit{subjective understanding} as described in previous research \cite{Rong2024,Knijnenburg2011,Radensky2022}. We present minority scenarios, i.e. group scenarios in which user preferences are similar except for one single member, a complex and potentially unfair scenario.

Our paper makes the following contributions:
\begin{itemize}[leftmargin=*]
    \item We conduct a preregistered, randomized controlled trail ($n=271$) to analyze the effect of different explanation types on user understanding.
    \item We show that more detailed explanations do not positively or negatively impact user understanding.
    \item We outline that the choice of aggregation strategy can impact understandability down the line.
    \item We find value in measuring understandability using a variety of tasks as opposed to a single metric.
\end{itemize}

The remainder of the paper is structured as follows. First, we outline the literature on GRS and understandability of Recommender Systems. Based on this body of previous work, we formulate our hypotheses. Second, we present our materials, including the measurement of understandability, explanation types, and experimental set-up. Subsequently, we describe the results of our experiment followed by a discussion of the results and their implications.

\section{Background}\label{sec:background}
In this section, we introduce the literature on group recommendation and social choice-based aggregation strategies. Additionally, we outline recent work combining understandability and Explainable AI (XAI) with Recommender Systems and highlight the research gap related to the understandability of GRS in particular.

\subsection{Group Recommendation}\label{sec:GRS}

\begin{table*}
  \caption{Social choice-based aggregation strategies derived from \cite{tran2019towards,barile2023evaluating,felfernig2018explanations}}
  \label{tab:strats}
  \begin{tabular}{ccl}
    \toprule
    Strategy & Type& Procedure \\
    \midrule
    Additive Utilitarion (ADD) & Consensus & Recommends the item with the highest sum of all group members’ ratings \\
    Fairness (FAI)& Consensus & Ranking and recommending items according to how individuals choose them in turn\\
    \midrule
    Approval Voting (APP) & Majority &  Recommends the item with the highest number of ratings above a predefined threshold\\
    \midrule
    Least Misery (LMS) & Borderline & Recommends the item which has the highest of all lowest per-item
ratings\\
    Most Pleasure (MPL) & Borderline & Recommends the item with the
highest individual group member rating\\
    
    \bottomrule
  \end{tabular}
\end{table*}

Recommendation systems tailored to groups, rather than the preferences of a single user, are rising in demand due to applications in fields such as tourism \cite{chen2021attentive} and music \cite{najafian2018generating}. These Group Recommender Systems (GRS) are designed to support the decision-making process of multiple people simultaneously \cite{Masthoff2022group}. The preferences of the individual group members need to be processed in a way that the recommendation reflects the group as a whole. Based on \textit{Social Choice Theory} \cite{kelly2013social}, social choice-based aggregation strategies are employed for aggregating the preferences of the group members to obtain a recommendation for the group as a whole \cite{masthoff2015group}. Examples of often used social-choice based approaches are summarized in Table \ref{tab:strats}.
A range of different strategies have been introduced in the literature and can be categorized as follows: \textit{consensus-based}, \textit{borderline} and, \textit{majority-based}. The latter makes use of only the most popular items or ratings \cite{senot2010analysis}. An example of such a strategy is Approval Voting (APP), which uses a threshold to recommend items having the higher number of ratings above a certain value (Table \ref{tab:strats}). 
Borderline strategies do not include all ratings, and filter out a subset of the item ratings in the group \cite{senot2010analysis}. For example, the Least Misery (LMS) strategy recommends the item with the highest of all lowest per-item ratings, while the Most Pleasure (MPL) strategy looks for the highest overall rating (Table \ref{tab:strats}). On the other hand, consensus-based strategies include all ratings made by the group to derive a group recommendation \cite{senot2010analysis}. An example of such a strategy is Additive Utilitarian (ADD). This strategy sums up all ratings per item and recommends the item with the highest sum (Table \ref{tab:strats}). In the current work, we make use of group recommendations derived by using a selection of strategies based on the categorization  presented by \citet{senot2010analysis}.

\subsection{Understandability of Recommender Systems}\label{sec:understanding}

We put social choice-based aggregation strategies from group recommendation in the frame of Human-Centered Explainable Artificial Intelligence (HCXAI) \cite{Rong2024,Ehsan2020}. It emphasizes the importance of human stakeholders regarding the desired goals of intelligent systems, focusing on factors including trust, usability, human-AI collaboration, and understandability \cite{Rong2024}. Specifically, understandability has attracted particular interest over the last few years \cite{Rong2024}. \citet{Liao2021} pointed out that understanding in the context of interacting with a machine learning (ML) model refers to the user's grasp or mental model of how the ML model operates. This knowledge grows from using the system and from receiving clear explanations about it. More precisely, the interpretability of an AI model should be defined according to its influence over the users' abilities in the completion of diverse tasks \cite{Doshi2017,Poursabzi2021}.

\citet{Knijnenburg2011} explored different ways in which users interact with an attribute-based recommender system. 
A post-experimental questionnaire is used to measure perceived understandability (using a 5-point Likert scale) of the interface. 
Increased domain knowledge resulted in higher understandability. 

\citet{Schroder2021} worked towards designing an understandable algorithmic experience in the music domain. 
Their small research-through-design experiment suggested that users comprehend the recommendations better when there is an easy path to accessing and understanding, and when the users are allowed to correct the system. Increased understanding could avoid a frustrated early majority.

\citet{Radensky2022} analyzed how providing local, global, or explanations incorporating both, influenced user understanding of the system behavior. The results indicated that the combination of both local and global explanation types are more helpful for explaining how to improve recommendations. However, the global explanations performed better for efficiently identifying false positive and negative recommendations of the users.

\citet{Guesmi2024} explored interactive explanations with a varying degree of detail using the intelligibility type categorization \cite{Lim2009}. The authors included dimensions such as \textit{What, What if, Why,} and \textit{How} and conducted a user study to investigate the impact on the perception of users while presented with interactive explanations. They concluded that, while allowing users to personalize the explanation type was an integral feature, the more advanced explanation types achieved the highest levels of transparency and trust \cite{Guesmi2024}.

Our measurement of understandability is based on the distinction between objective and subjective understanding. \citet{Rong2024} pointed out that measuring objective understanding is usually done by deploying proxy tasks to verify user comprehension of a certain computational model. Various methodologies have been explored in the literature and serve as foundation of the current work, including simulating model predictions~\cite{Cheng2019,Wang2021}, evaluating counterfactual thinking~\cite{Hohman2019,Wang2021}, and detecting mistakes~\cite{Wang2021}. \citet{Wang2021} showed that counterfactuals and visualizations of feature importance increased objective understanding. 
\citet{Cheng2019} illustrated that white-box models increased the ability to simulate model behavior and concluded that interactivity is an important factor when it comes to improving objective understanding.

On the other hand, subjective understanding is usually measured by post-task questionnaires \cite{Knijnenburg2011, Radensky2022}. These self-assessments are performed by Likert scale questions and statements such as ``\textit{I understand the decision algorithm}'' and ``\textit{The explanation helps me to understand}''. Subjective understanding has been tested based on different implementations of explanations such as rule-based explanations \cite{Buccinca2020}, example- and counterfactual-based scenarios \cite{Wang2021}, or LIME and SHAP explanations \cite{Hadash2022}. The meta-analysis by \citet{Rong2024} outlined that explanations of a computational model increased subjective understanding of users. 

\subsection{Social Choice-based Explanations}\label{sec:grs-exp} 
In the context of GRS, explanations have been used to assess a variety of factors such as consensus perception~\cite{barile2023evaluating,felfernig2018explanations} or privacy-preservation~\cite{najafian2021exploring,najafian2021factors}. Such explanations are natural language excerpts, typically outlining the underlying mechanism of the social choice-based aggregation strategy \cite{najafian2018generating,kapcak2018tourexplain}. User studies evaluated such explanations in terms of fairness perception, consensus perception and satisfaction with mixed results~\cite{tran2019towards,barile2023evaluating}. Our control condition (no explanation) as well as the textual explanations follow the template presented in prior work. 

\subsection{Hypotheses}\label{sec:hypotheses}
All in all, understandability of recommender systems and ML models in general has received ample attention. Additionally, explanations for GRS have been discussed and evaluated on the basis of a diverse list of factors. However, the concept of understandability of social choice-based explanations remains fairly unexplored. 
In this work, we aim to address this research gap by implementing the concept of understandability based on a varying degree of social choice-based explanations. 

In light of the literature described in Section \ref{sec:understanding}, we hypothesize that both objective and subjective understanding will increase when the user is presented with an explanation, with a bigger effect when presented with a multimodal explanation. More formally, we define the following \textit{hypotheses related to RQ1}:\newline

\noindent \textbf{H1a}: Explanations will lead to a higher level of objective understanding of the underlying aggregation strategy, with a bigger effect for multimodal explanations.\newline

\noindent \textbf{H1b}: Explanations will lead to a higher level of subjective understanding of the underlying aggregation strategy, with a bigger effect for multimodal explanations.\newline

In Section \ref{sec:grs-exp}, we discussed the mixed results related to the inclusion of textual social choice-based explanations on factors such as fairness or satisfaction \cite{tran2019towards, barile2023evaluating}. However, these user studies did find differences between strategies themselves. Based on these divergent outcomes among strategies, we formulate the following hypotheses:\newline

 \noindent \textbf{H2a}: The effect of the explanations on the level of objective understanding is moderated by the underlying social choice-based aggregation strategy used to derive the recommendation. \newline

 \noindent \textbf{H2b}: The effect of the explanations on the level of subjective understanding is moderated by the underlying social choice-based aggregation strategy used to derive the recommendation.

\section{Methodology}
In this section, we present the methodology and procedure for our user study. Our experiment was approved by the ethical committee of Maastricht University\footnote{\url{https://www.maastrichtuniversity.nl/ethical-review-committee-inner-city-faculties-ercic}}. The between-subject design, including research questions, variables and hypotheses, was preregistered on Open Science Framework (OSF).\footnote{\url{https://osf.io/myx7p/?view_only=597bf08540a94dbd864b420ce3351a7d}}

\subsection{Materials}\label{sec:material}
\begin{table*}[h]
  \caption{Textual explanation presented to participants in both \textit{text\_expl} and \textit{graph\_expl} groups}
  \label{tab:textexp}
  \begin{tabular}{cl}
    \toprule
    Strategy & Textual explanation\\
    \midrule
    ADD &  $i_k$ has been recommended to the group since it achieves the highest total rating.\\
    LMS & $i_k$ has been recommended to the group since no group members has a real problem with it.\\
    APP & $i_k$ has been recommended to the group since it achieves the highest number of ratings which are above 3.\\

    \bottomrule
  \end{tabular}
\end{table*}

\subsubsection{Group scenarios}

Inspired by \citet{barile2023evaluating}, our study makes use of a series of scenarios, each containing a hypothetical group which is being recommended a restaurant. 
Each scenario consists of a table in which a group of five members is presented alongside their fictitious ratings of 10 restaurants (on a scale from 1 to 5). Such groups can be built using distinct configurations, representing differing degrees of agreement among group members' preferences. Configurations outlined in previous work are \textit{divergent} (high diversity), \textit{uniform} (low diversity), \textit{coalitional} (two distinct sub-groups) and \textit{minority} (low diversity with the exception of one member) \cite{barile2023evaluating}. While a \textit{uniform} configuration presents a simplistic scenario which needs no explanation, \textit{divergent} and \textit{coalitional} configurations lead to difficulty finding recommendations that satisfy group members. Thus, in this study, we strictly constructed fictitious groups based on the \textit{minority} configuration. Each scenario was generated using the computational procedure outlined by \citet{barile2023evaluating}. However, since that the presented scenario considers a group who already used the system three times and receives a fourth recommendation (see Section \ref{sec:procedure}), we also imposed the constraint that each scenario generated for a specific strategy would not have ties (items with the same group score) at the fourth interaction. Finally, the scenario is completed with anonymous items (named \textit{$Rest_i$}), and random names for the group members (from a list of gender neutral names, to minimize the risk of possible biases).\footnote{The source code for generating the group configurations and explanations is available at the following link: \url{https://anonymous.4open.science/r/Understanding_GRS-1E7F}}

\subsubsection{Aggregation strategies}
Participants were assigned one of three social choice-based aggregation strategy and were only presented with scenarios in which that strategy was used to derive a group recommendation. To ensure a varied selection of strategies to derive a group recommendation, we included one of each category presented in Table \ref{tab:strats}. Practically speaking, each participant was shown recommendations made by either the Additive Utilitarian (ADD), Least Misery (LMS) or Approval Voting (APP) strategy, respectively covering consensus-based, borderline and majority-based aggregation categories (Table \ref{tab:strats}). The threshold rating for the APP strategy was set at $3$, equal to 60\% of the rating scale.

\subsubsection{Explanations}
\begin{figure*}[h]
  \centering
  \includegraphics[width=.7\linewidth]{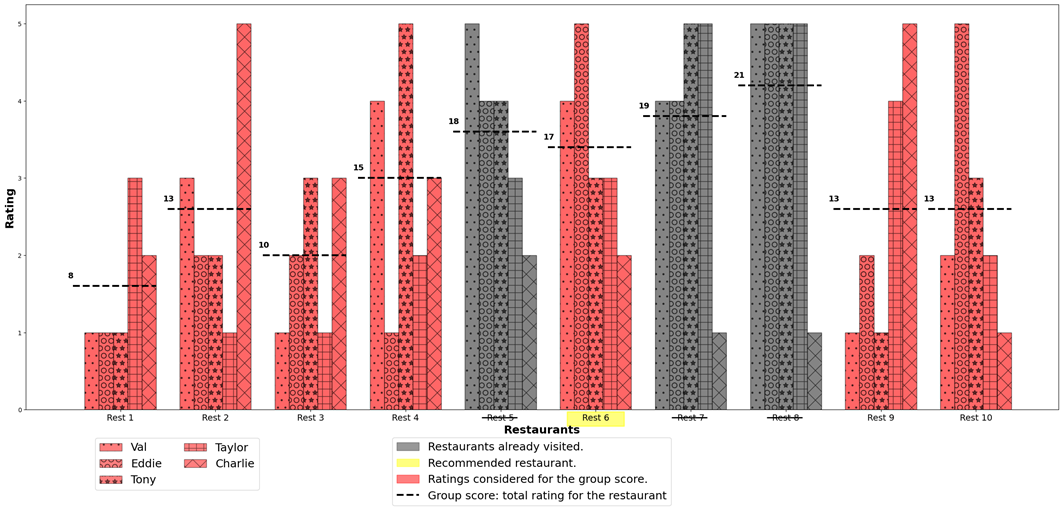}
  \caption{Graphical explanation for the Additive Utilitarian (ADD) strategy (example figure). Already visited restaurants are shown in grey. Potential recommendations are colored red. The output of the system is highlighted in yellow. Bar charts for ADD include a horizontal line indicating the sum of all per-item ratings.}
  \label{fig:graphexp}
  \Description{Example figure of graphical explanation. The example shows the graphic of a scenario using the Additive Utilitarian aggregation strategy (ADD). Already visited restaurants are colored grey, while potential candidates are red. For each restaurant, a horizontal line indicates the sum of all ratings for that specific restaurant. The chosen restaurant by the system is colored yellow on the x-axis.}
\end{figure*}

Alongside a strategy, each participant was assigned one explanation type. In total, this study included three explanation types, (two explanation types supplemented with a control condition). The first explanation level (\textit{no\_expl)}, the control condition, included no explanation. The participant was only presented with the group ratings and the output. The second level (\textit{text\_expl}) provided a simple textual social choice-based explanation, adopted from previous work \cite{barile2023evaluating} and summarized in Table \ref{tab:textexp}. The third and final explanation type supplemented the textual explanation with a graphical representation of the procedure using a bar chart, an often used visualization for traditional user interfaces crafted for explaining recommendations \cite{Gedikli2014, Tintarev2015}. This multimodal explanation (\textit{Graph\_expl}) visualized the ratings of each group member, as well as the already chosen restaurants and the current recommendation (Figure \ref{fig:graphexp}). The ratings influencing the specific outcome were colored red. A final component differed among strategies. The ADD strategy included a line indicating the sum of all ratings per restaurant (Figure \ref{fig:graphexp}), while graphics for the LMS strategy showcased a line corresponding to the lowest per-item rating. Finally, graphics visualizing the APP strategy included a line indicating how many ratings are above the set threshold. 

\subsection{Variables}
Using the materials described in Section \ref{sec:material}, our study consisted of two independent, between-subject variables.

\begin{itemize}[leftmargin=*]
    \item \textbf{Exp} (categorical, between-subject): each participant is randomly assigned an explanation modality:
    
    $Exp_{i} \in [no\_expl, text\_expl, graph\_expl]$;
    \item \textbf{Agg} (categorical, between-subject): each participant is randomly assigned one social choice-based aggregation strategy: 
    
    $Agg_{i} \in [ADD, LMS, APP]$;
\end{itemize}


Additionally, our experiment included two main dependent variables:\textit{ objective} and \textit{subjective} understanding. Motivated by \citet{Wang2021}, we designed a set of questions based on three aspects:
\begin{itemize}[leftmargin=*]
    \item Simulate model behavior: \textit{Giving a new scenario, choose the right recommendation.}
    \item Counterfactual thinking: \textit{Giving a new scenario and a given recommendation, pick the answer that results in an alteration of that output.}
    \item Error detection: \textit{Giving a new scenario and a given recommendation, identify whether the presented recommendation is correct.}
\end{itemize}

For each participant, two tasks for each of these three aspects were presented. Each task was given a 0/1 value (0 = incorrect, 1 = correct). Per participant, these scores were summed up and divided by the number of tasks to obtain their accuracy rate:

\begin{itemize}[leftmargin=*]
    \item \textbf{Objective understanding} (continuous): The accuracy rate of each participant's answers to the objective understanding questionnaire (six questions in total). 
\end{itemize}

For measuring subjective understanding, participants were asked to rate two statements previously used by \citet{Wang2021}:
\begin{itemize}[leftmargin=*]
    \item \textit{``I understand how the model works to predict the best recommendation for the group.''}
    \item \textit{``I can predict how the model will behave.''}
\end{itemize}

These questions were presented two times, once at the end of the training step, and finally at the end of the experiment. From these, we compute the following variables:

\begin{itemize}[leftmargin=*]
    \item \textbf{Preliminary subjective understanding} (continuous): The average between both subjective understanding statements asked \textbf{before} the objective understanding measurement.
    \item \textbf{Final subjective understanding} (continuous): The average between both subjective understanding statements asked \textbf{after} the objective understanding measurement
\end{itemize}

In the analysis described in Section \ref{sec:results}, we strictly made use of \textit{Final subjective understanding}, measured at the end of the survey.

In addition to the independent and dependent variables that we used for hypothesis testing, we collected age group and gender data to allow for a demographic description of our sample. 

\begin{itemize}[leftmargin=*]
    \item \textbf{Age} (categorical). Participants will be able to select one of the options \textit{$<$18}, \textit{18-25}, \textit{26-35}, \textit{36-45}, \textit{46-55}, \textit{$>$55}, or \textit{Prefer not to share this information}.
    \item \textbf{Gender} (categorical). Participants will be able to select one of the options \textit{Female}, \textit{Male}, \textit{Nonbinary}, \textit{A gender not listed here}, or \textit{Prefer not to share this information}.
\end{itemize}

\subsection{Procedure}
\label{sec:procedure}

\begin{figure*}[h]
  \centering
  \includegraphics[width=.9\linewidth]{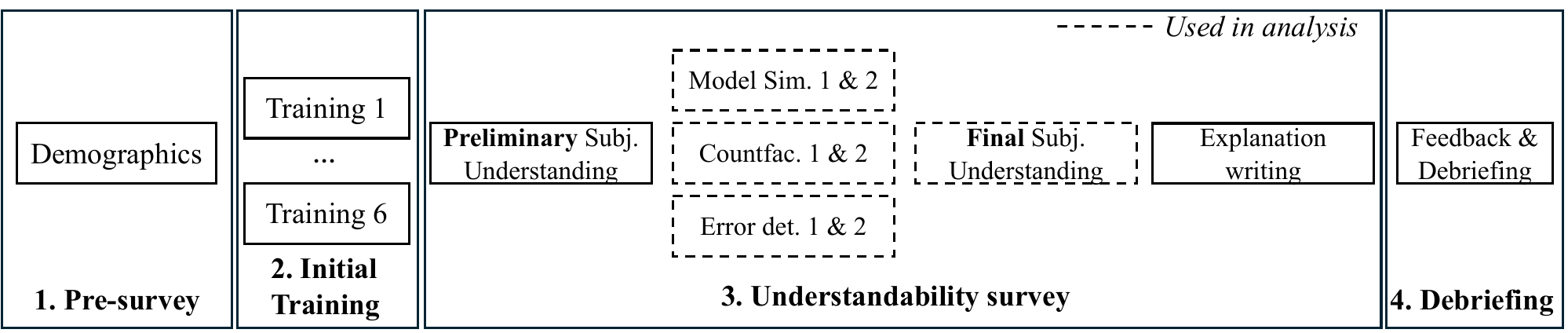}
  \caption{Survey components in the order as seen by participants. Survey questions in dotted lines were used to derive the dependent variables.}
  \label{fig:pipeline}
  \Description{Visual representation of the survey components and their order. First, a participant filled in their demographic. Second, six training scenarios were presented which showcased the explanation category they were assigned to. Subsequently, each participant had to fill in the understandability survey which contained subjective statements, six objective understandability tasks, the subjective statements again and finally, a prompt to write an explanation themselves. Last, participants were given the opportunity to provide feedback and were shown a debriefing message.}
\end{figure*}

The survey was executing with the Qualtrics tool\footnote{\url{https://www.qualtrics.com/}} and consisted of four stages (Figure \ref{fig:pipeline}).

\subsubsection*{Pre-survey} Participants were randomly assigned to one of the three social choice-based aggregation strategies and to one of the three explanation types. First, they agreed to an informed consent, and indicated their age group and gender. Afterwards, instructions were given detailing the procedure of the training phase.

\subsubsection*{Initial training} Each participant was presented with a series of six scenarios, consisting of one hypothetical group recommendation and configured on their assigned aggregation strategy and explanation type. An attention check was included in the fourth training scenario. No dependent variables are recorded during training. For each scenario, the participant followed a three-step procedure:

\begin{enumerate}
    \item Make an initial recommendation decision regarding the restaurant recommendation.
    \item Review the recommendation delivered by their assigned social choice-based strategy and, possibly, the explanation.
    \item Make a final recommendation.
\end{enumerate}

Each scenario was introduced by the following excerpt: 
\textit{Assume that there is a group of friends. Every month, a group decision is made by these friends to decide on a restaurant to have dinner together. To select a restaurant for the dinner next month, the group again has to take the same decision. In this decision, each group member explicitly rated ten possible restaurants using a 5-star rating scale (1: the worst, 5: the best). The ratings given by group members are shown in the table below. Under the table, you also find the order of restaurants the group has already visited in the previous months. These restaurants are not an option anymore, as the group has already eaten there previously.}

\subsubsection*{Understandability survey} After training, each participant was presented with the two subjective understanding statements. After rating their self-perceived understanding, six questions for testing objective understanding needed to be answered. Two questions are aimed at model simulation, two measured counterfactual thinking and finally, two questions geared towards error detection\footnote{Full documentation including all questions, answer options and visualizations can be found in the OSF folder: \url{https://osf.io/myx7p/?view_only=597bf08540a94dbd864b420ce3351a7d}}. 
Additionally, a second attention check was presented on the page of the first counterfactual scenario. After the tasks geared towards objective understanding, the two subjective statements were presented again. For the final calculation of subjective understanding, we made use of these final two ratings. This part of the experiment ended with a final task: the participant are presented with the first training scenario, and asked to provide an explanation for the group, in their own words, on how the system derived a group recommendation. The responses gathered during this survey were used to derive the dependent variables.

\subsubsection*{Debriefing} The participants had the option to provide additional feedback in an open text field. Finally, a short debriefing message was showed, before redirecting them to the recruitment platform.

\subsection{Sample Size Determination}

We computed the minimum required sample size in a power analysis for the planned ANOVA tests (see Section \ref{sec:results-RQs}) using \textit{G*Power}~\cite{GPowerFaul2007}. To account for multiple hypothesis testing, we applied a Bonferroni correction by adjusting our significance threshold $\alpha = \frac{0.05}{2} = 0.025$. We calculated the sample size for an ANOVA (Fixed effect, special, main effects and interaction). We specified an effect size f = 0.25, $\alpha  = 0.025$, a power of $(1-\beta) = 0.85$, degrees of freedom numerator equal to 4, and $3 \times 3 = 9$ groups (i.e., $3$ different aggregation strategies for $3$ different explanation scenarios). These specifications aligned with previous works \cite{Cheng2019,Wang2021}. These calculations resulted in a minimum required sample size of $257$ participants. The analysis only included the \textbf{final} subjective understanding (one measurement per participant). As a result, we did not use a repeated measures ANOVA.

\begin{figure*}[ht]
  \centering
  \includegraphics[width=1\linewidth]{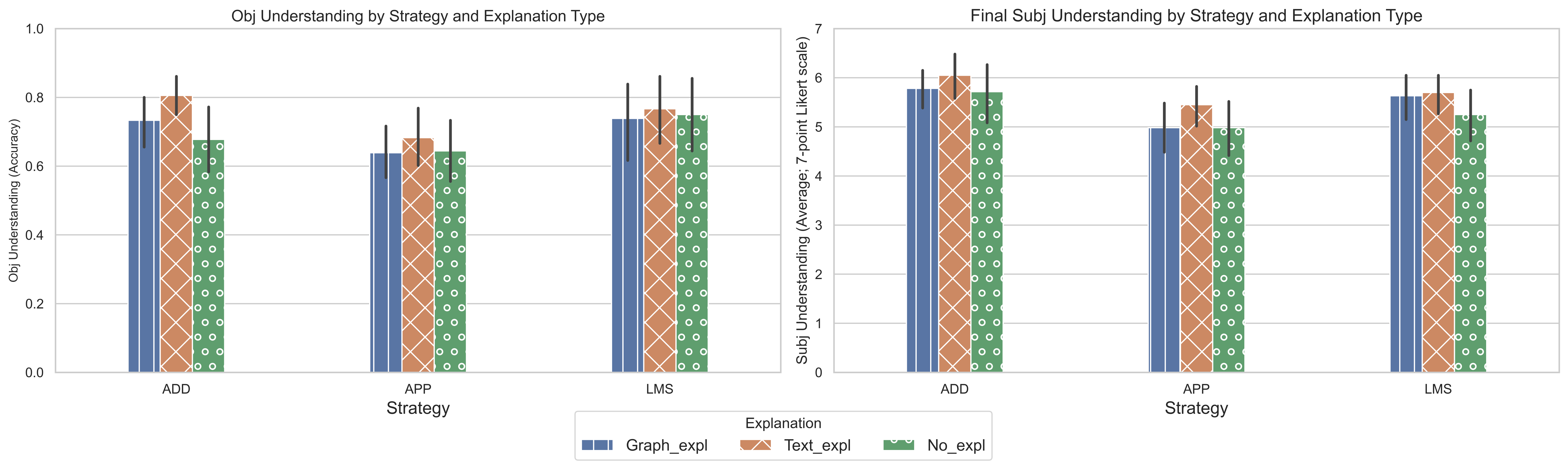}
  \caption{Objective (left, 0-1 scale) and subjective (right, 0-7 scale) understanding clustered by aggregation strategy (ADD = additive utilitarian, APP = approval voting, LMS = Least misery) and explanation modality (no\_expl = control group, text\_expl = textual, Graph\_expl = multimodal)}
  \label{fig:summary-obj-subj}
  \Description{Bar chart summarizing the objective and subjective understanding. Clustered by aggregation strategies: ADD = Additive utilitarian, APP = Approval voting, LMS = Least misery. Left chart shows objective understanding calculated as the accuracy rate of answer given to six tasks (0 to 1). The right plot shows subjective understanding calculated as the average of two 7-point Likert scale questions. The blue bar (left of three) is the control group (no\_expl), the orange bar (middle) is the textual explanation group (text\_expl)) and the green bar (right) is the multimodal group (graph\_expl)}
\end{figure*}

\section{Results}\label{sec:results}
In the following section, we report the results from our experiment. The full data is linked in the OSF folder\footnote{\url{https://osf.io/myx7p/?view_only=597bf08540a94dbd864b420ce3351a7d}}. We describe our sample, and present the overall understandability of the task, that helps us provide an indication of the complexity of the scenarios and illustrates the relationship between objective and subjective understanding. Afterwards, we address the research questions. Finally, we perform exploratory analyses regarding the different aggregation strategies and objective understanding tasks.

\subsection{Participants}
We recruited a total of $290$ participants using the online participant pool \textit{Prolific}.\footnote{\url{https://prolific.co}} They were required to be proficient English speakers above 18 years of age. Each participant was allowed to participate in the study once, and received a reimbursement according to Prolific guidelines\footnote{\url{https://www.prolific.com/resources/how-much-should-you-pay-research-participants}}, considering a hourly reward of £9. After removed $19$ participants, due to failed attention checks, our final sample consisted of $271$ participants: 
$30$ participants for each of the nine conditions (characterized by one strategy and one explanation type), with the exception of the \textit{APP text\_expl} group ($n = 31$). Our sample was composed of 51\% (139) female, 48\% (131) male and 1\% (1) nonbinary participants. Additionally, 42\% (115) were between 26 and 35 years old, 30\% (80) between 18 and 25, 16\% (44) were between 36 and 45 years old, 7\% (18) between 46 and 55 and 5\% (14) indicated they were older than 56 years old.

\subsection{Overall Understandability}
Average understanding scores are visualized (by strategy and explanation type) in Figure \ref{fig:summary-obj-subj}. For objective understanding, we show accuracy scores which were calculated based on all six tasks. For subjective understanding, we the objective understanding survey (\textit{Final subjective understanding}), which was recorded \textbf{after} the objective understanding tasks. Subjective understanding was not treated as repeated measure; \textit{preliminary subjective understanding} was \textbf{not} used in this analysis. Overall, understandability was relatively high. The average objective understanding was $0.72$ ($SD=0.25$) on a scale of 0 to 1, while the average (final) subjective understanding (scale from 0 to 7) was $5.51$ ($SD = 1.36$).  

\begin{table}[h]
  \caption{Results of two two-way ANOVAs for dependent variables (DV) objective and subjective understanding. The independent variables were explanation type (Exp) and aggregation strategy (Agg)}
  \label{tab:anova}
  \begin{tabular}{lllll}
    \toprule
     & \multicolumn{2}{c}{DV: Objective} & \multicolumn{2}{c}{DV: Subjective}\\
    \midrule
    Group & \textit{F}& \textit{p} & \textit{F}& \textit{p}\\
    \midrule
    Exp & 1.490 & 0.227 & 2.274 & 0.105\\
    Agg & 4.081 & 0.018* & 6.350 & 0.002**\\
    Exp:Agg & 0.438 & 0.781 & 0.246 & 0.912\\
    \bottomrule
  \end{tabular}
\end{table}

\subsection{Hypotheses Validation}\label{sec:results-RQs}
Regarding the \emph{``RQ1: Do explanations increase users’ understanding?''} we found no significant differences between the three explanation types, for both objective understanding (\textbf{H1a;} $F=1.49, p=0.23$ -- see Table \ref{tab:anova}) and subjective understanding (\textbf{H1b;} $F=2.27, p=0.10$ -- see Table \ref{tab:anova}). Thus, increasing detail, whether textual or multimodal, did not have an impact on user understanding as we did not find a significant difference compared to the control group (no explanation).
Focusing on the \emph{RQ2: ``Does the effect of explanations vary depending on the aggregation strategy?''}, we also did not find significant interaction effects between aggregation strategies and explanation types regarding objective (\textbf{H2a;} $F=0.44, p=0.78$  -- see Table \ref{tab:anova}) and subjective understanding (\textbf{H2b;} $F=0.25, p=0.91$ -- see Table \ref{tab:anova}).

\subsection{Exploratory Analysis}\label{sec:results-exploratory}
Additional to testing our hypotheses, we performed some exploratory analyses to understand the impact of social choice-based aggregation strategies included in the experiment. Additionally, we attempted to explain some of the variance in objective understanding.
As shown in Table \ref{tab:anova}, our analysis shows significant differences between aggregation strategies. Tukey pairwise post-hoc analysis revealed that participants assigned to LMS achieved higher objective understanding ($p_{adj}=0.024$) compared to the participants assigned to APP. Additionally, participants exposed to the ADD strategy indicated a higher subjective understanding compared to those assigned to APP ($p_{adj}=0.001$). 
Finally, we analyzed the different tasks making up our measurement of objective understanding: \textit{model simulation}, \textit{counterfactuals} and \textit{error detection} (see Table \ref{tab:obj-tasks}). Overall, counterfactuals resulted in a lower average accuracy, implying a rather difficult task. Explanations, however, did not increase average counterfactual accuracy rates.
On the other hand, error detection assignments resulted in a relatively high accuracy across the board. 

\begin{table}[h]
  \caption{Average accuracy (and standard deviation) for objective understanding tasks; range between 0 and 1}
  \label{tab:obj-tasks}
  \begin{tabular}{lccc}
    \toprule
    & Model sim. & Counterfactual & Error detection \\
    \midrule
    No\_expl & 0.73 (0.37) & 0.52 (0.43)& 0.82 (0.31) \\
    Text\_expl & 0.81 (0.31) & 0.54 (0.43)& 0.91 (0.21)\\
    Graph\_expl & 0.74 (0.39) & 0.53 (0.41)& 0.84 (0.25)\\
    \bottomrule
  \end{tabular}
\end{table}

\section{Discussion} 
In this study, we evaluated the effect of explanations for group recommendations on user understandability. Additionally, we looked at the outcomes based on the different social choice-based aggregation strategies used, and compared the accuracy rates of the three distinct assignment types making up our objective understanding measurement. In the following section, we discuss the results from our experiments and formulate potential causes. Additionally, we argue in favor of measuring understanding using diverse assignments and formulate the implications of our study.

\subsection{The Lack of Impact of Explanations}

We did not find any significant differences between the three explanation types included in our experiment. This was the case for both objective and subjective understandability (Table \ref{tab:anova}). We did not find a negative impact of more complex explanations, which was reported in \citet{Kaur2024}. This divergent outcome may be caused by the fact that the more complex explanation types analyzed by \citet{Kaur2024} involved interactive elements, while our more complex explanation category did not. However, we can observe a trend in which textual explanations (\textit{text\_expl}) achieved slightly higher understanding (both objective and subjective) compared to the more complex, multimodal explanation (Figure \ref{fig:summary-obj-subj}). 

Our results are in line with previous work analyzing overall effect of explanations in group recommendation scenarios: 
\citet{barile2023evaluating} did not find an impact of explanations on perceived consensus, perceived fairness and satisfaction. Furthermore, our outcomes are in line with a recent analysis on the understanding of machine learning models by \citet{Rong2024}, which reported mixed results on the impact of explanations on objective understanding. 


The lack of significant effects between the different explanation conditions may depend on several reasons. 
A first possible motivation might be related to the tasks, that could have been too hard or abstract. However, the participants' understanding is relatively high across the board, including the control group (\textit{no\_expl}) -- see Figure \ref{fig:summary-obj-subj}. Additionally, the total time spent by participants was higher than expected; if participants disengaged due to complexity, we would expect a lower accuracy and a shorter duration.

\begin{figure}[h]
  \centering
  \includegraphics[width=1\linewidth]{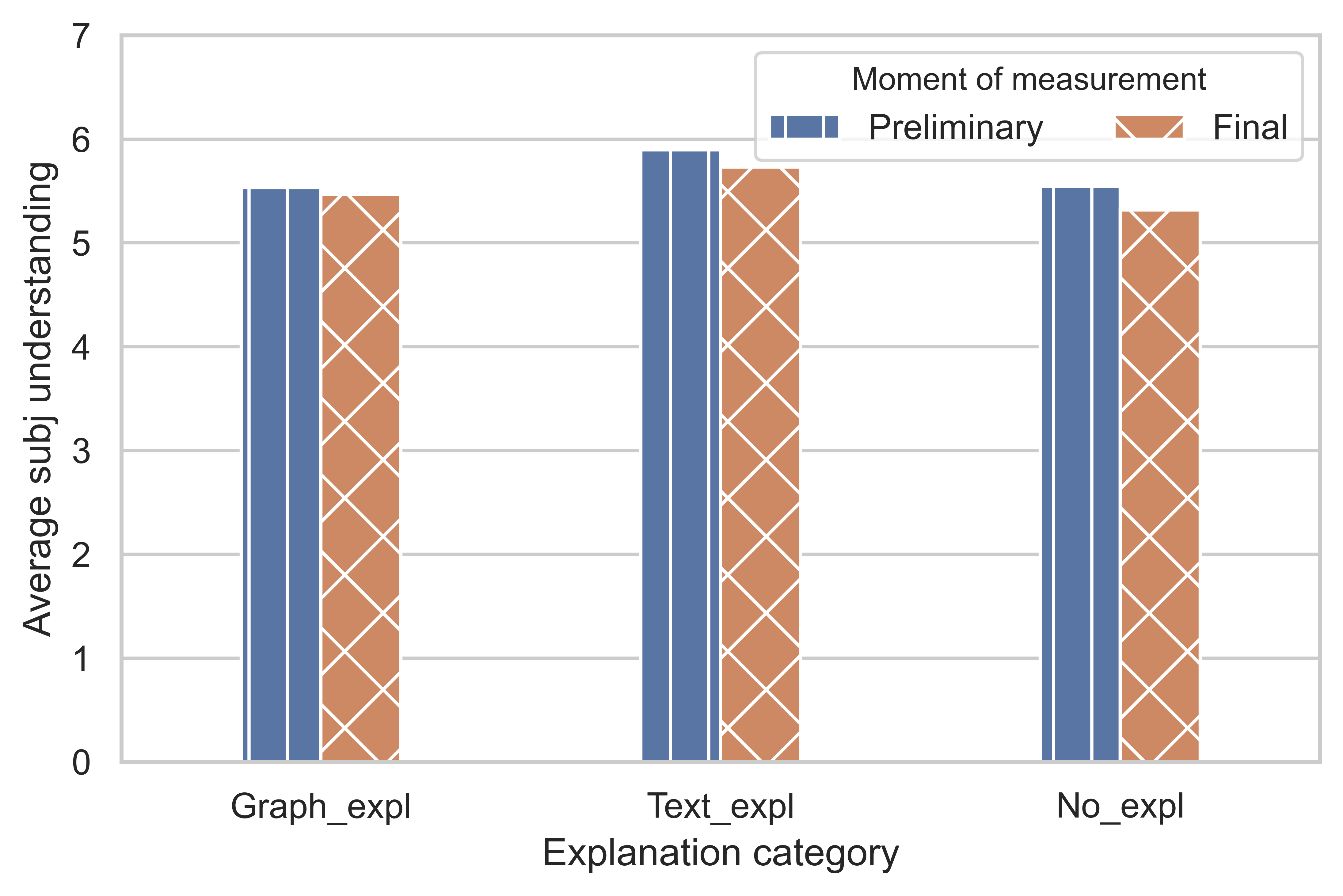}
  \caption{Subjective understanding (7-point Likert scale) measured before (preliminary) and after (final) the objective understanding tasks; by explanation modality (no\_expl = control group, text\_expl = textual, Graph\_expl = multimodal)}
  \label{fig:subj-before-after}
  \Description{Bar chart comparing subjective understanding measured before and after the objective understanding tasks. Subjective understanding is calculated as the average of two survey questions (7-point Likert scale, 0-7 range). Clustered by explanation modality (x-axis from left to right): control group (No\_expl), textual explanations and multimodal explanations). Blue bar (left bar) shows subjective understanding measured before, orange bar (right bar) shows subjective understanding after}
\end{figure}

On the other hand, the used scenarios might have been too easy, leading to high understandability regardless of the presence of an explanation. We argue against this argument by comparing our subjective understanding measurement used in the analysis with the \textit{preliminary subjective understanding}, measured between the training phase and objective understanding tasks: we observed a decrease in self-perceived understanding after participants completed the six objective understanding tasks (Figure \ref{fig:subj-before-after}). 
This trend, combined with the fact that we did not receive participant feedback that implied a lack of difficulty, indicates that the group recommendation scenarios might not have been too simple. 

A more likely motivation for our results can be found within \textit{bounded rationality}. As discussed by \citet{Kaur2024} in the context of interpreting machine learning models, humans are inclined to achieve \textit{good enough} understanding (\textit{satisficing}) as opposed to optimizing their decisions. In our study, it is possible that participants were satisficed with the group ratings and previously recommended restaurants, disregarding additional explanations. The relatively high understanding for the control group presents evidence in favor of bounded rationality (Figure \ref{fig:summary-obj-subj}). However, future work needs to unpack the individual degree of satisficing for group recommendation further, for example by asking participants to indicate which information they used to make their decision. Additionally, the spectrum of presented information, both within the scenario and explanation, can be expanded to pinpoint when participants are provided enough information for a satisficing outcome.

\subsection{Performance on Objective Understanding Tasks}
In Section \ref{sec:results-exploratory}, we separately presented the average accuracy rates for each of the three objective understanding tasks to investigate the variance within the objective understanding measurement. Error detection resulted in the highest overall accuracy (Table \ref{tab:obj-tasks}). However, these questions were binary (correct/incorrect), while the other two comprised four or more options. Additionally, we found that, compared to both model simulation and error detection, counterfactuals turned out to be a difficult task. A potential cause for this lower accuracy may be the fact that our counterfactual assignment required participants to perform multiple model simulations in sequence.\footnote{All scenarios and corresponding tasks are found in the OSF folder: \url{https://osf.io/myx7p/?view_only=597bf08540a94dbd864b420ce3351a7d}} This is reflected in the timestamp of the assignments. On average, it took participants longer to complete a counterfactual assignment compared to the other two tasks (Table \ref{tab:timing-obj-tasks}). This post-hoc analysis suggests the value of measuring objective understanding based on the performance on a variety of tasks, as different types of tasks lead to divergent results.

\begin{table}[h]
  \caption{Average time (and standard deviation) in seconds for completing a single assignment; per objective assignment task}
  \label{tab:timing-obj-tasks}
  \begin{tabular}{lccc}
    \toprule
    & Model sim. & Counterfactual & Error detection \\
    \midrule
    No\_expl & 51.5 (40.1) & 113.2 (85.9)& 36.8 (26.4) \\
    Text\_expl & 58.5 (114.2) & 114.3 (74.0)& 39.0 (25.5)\\
    Graph\_expl & 48.4 (35.7) & 111.6 (94.9)& 39.5 (32.2)\\
    \bottomrule
  \end{tabular}
\end{table}

\subsection{Implications}

Our findings have clear implications which need to be considered by system designers and other practitioners. First, the results presented in Section \ref{sec:results-exploratory} show that the social choice-based aggregation strategy implemented within a GRS may influence user understandability down the line. When designing a GRS, one should be mindful that these methodological decisions should be weighed not only in terms of fairness, consensus and satisfaction, but also in terms of user understanding. If user understanding and explainability are primary concerns, opting for a more straightforward aggregation strategy in the design phase is beneficial. 

Second, our experiment provides an indication to system designers looking to improve user understandability by adapting certain elements in the interface. Against intuition, textual explanations were not useful in improving user understanding. Similarly, adding visual elements to the textual explanation did not increase user understanding either. Thus, practitioners need to look towards other procedures (other than explanations) to improve user understanding of GRS. The relatively high understanding achieved by the control group implies that the presentation of previous output alongside group context might already provide sufficient information to users. 

All in all, our results imply that increased detail in explanations might become redundant when users receive a satisficing amount of information already. This information and explainability can be derived from the implemented methodology (e.g. strategy) and context clues such as group ratings and previous output. System designers and other practitioners should keep in mind that simply increasing the details presented in explanations will not necessarily translate into improved user understanding. For GRS specifically, opting for social choice-based aggregation strategies such as \textit{Least Misery} or \textit{Additive Utilitarian}, as opposed to \textit{Approval Voting}, might be beneficial to ensure user understanding. 

\subsection{Limitations}
We identified several limitations that may have had an impact on our results. First, we focused on the minority group configuration; this limits the generalizability of our results. As discussed in \citet{barile2023evaluating}, different group configurations may lead to different evaluations in terms of satisfaction and perception of fairness and consensus. Future works are necessary to evaluate the possibility of different impact of explanations for different group configuration.

Additionally, we did not ask participants to evaluate their AI literacy. \citet{Kaur2024} highlighted that Machine Learning practitioners were faster but less accurate when having access to interpretability tools. 
Future research could include self-rated measurements of AI literacy and compare practitioners and novices in terms of understandability of GRS. 

Finally, we presented group scenarios involving the recommendation of unnamed restaurants. However, high investment domains such as tourism could have a higher need for explainability. Future work could make the comparison between low and high investment domains in group scenarios to investigate whether domain-specific factors influence the impact of explanations.

\section{Conclusion}
In this study, we presented a randomized controlled trial to analyze whether increasing detail of social choice-based explanations for GRS improved understandability. Our setup consisted of two between-subject factors: (i) the explanation modality, and (ii) the social choice-based aggregation strategy used to generate the group recommendations. We constructed two ANOVA models using objective and subjective understanding as dependent variables. However, we did not find significant differences between explanation types (no explanation, textual and multimodal explanation). Hence we outlined potential causes for this result rooted in bounded rationality: the group context and previous recommendations may suffice to provide participants with a satisficing level of understanding.

Furthermore, we conducted several post-hoc analyses to explain some of the variance found within our understanding measurements. We found that the methodological choice regarding aggregation strategy can impact understandability down the line. Additionally, our results indicate that counterfactuals tasks were more difficult compared to model simulation and error detection. We conclude that there is value in measuring understanding based on a multitude of assignments. 

Finally, we discussed some implications of our work. Besides factors such as fairness and satisfaction, the methodological choice of aggregation strategy needs to be weighed in terms of understandability. Additionally, system designers looking to adapt user interfaces 
need to mindful of the fact that additional elements do not necessarily translate to improved understandability.


\begin{acks}
This work was supported by the Faculty of Science and Engineering at Maastricht University (2024 FSE Starter Grant).
R. Yera Toledo is supported by the European Union's Horizon Europe research and innovation program under the Marie Sklodowska-Curie grant agreement number 101106164.
\end{acks}

\bibliographystyle{ACM-Reference-Format}
\bibliography{reprobib}


\end{document}